\documentclass[sort&compress
    ,final            
    ,numbers              
  ]
  {aipproc}

\usepackage{amssymb,amsmath,amstext,amsthm,amsfonts}
\layoutstyle{8x11single}

\usepackage[usenames]{color}
\definecolor{magenta}{cmyk}{0,1,0,0}

\newcommand{\plotscale}{.65}

\begin{document}

\title{Neutrino-Nucleus Interactions around 1 GeV}


\author{Ulrich Mosel}{
  address={Institut fuer Theoretische Physik, Universitaet Giessen, D-35392 Giessen, Germany},
  email={mosel@physik.uni-giessen.de},
  thanks={Work supported by DFG and HIC for FAIR}
}

\begin{abstract}
The extraction of neutrino oscillation parameters often requires the unambiguous identification of the reaction mechanism to be used for the energy determination. The use of nuclear targets and of beams with broad energy distributions complicates this task because it requires the reliable description of many different elementary processes and the use of event generators to isolate and identify them. In this talk I briefly summarize our present understanding of this task.

\end{abstract}

\maketitle

The extraction of neutrino oscillation parameters often requires the unambiguous identification of quasielastic (QE) scattering to be used for the neutrino energy determination. The use of nuclear targets and of beams with broad energy distributions complicates this task because it requires the reliable description of many different elementary processes and the use of event generators to isolate and identify them. In particular, in the quasielastic regime additional, overlapping contributions could come from both pion production and many-particle interactions.

The close entanglement of the pion-production channel (mainly through the $\Delta$ resonance) with the quasielastic one has recently been reviewed in \cite{Leitner:2010kp,Mosel:2011qx}. There it has been shown that both tracking and Cherenkov detectors require large corrections for the cross section determination that even depend quite strongly on experimental acceptance cuts. It was also shown that Cherenkov detectors necessarily shift the reconstructed energy towards smaller energies.

The MiniBooNE has observed that the cross section identified as being that of quasielastic scattering is significantly larger than the QE cross section obtained by various different theories and generators when they all use the same, world-average axial mass $M_A \approx 1 $ GeV (see Fig.\ 15 in \cite{AguilarArevalo:2010zc}). This has triggered many discussions and even a call for a new paradigm \cite{Benhar:2010nx}.  Possibly connected with this surplus in QE scattering is the very large pion production cross section obtained in that experiment \cite{Leitner:2009de}. The measured pion yield is significantly larger than theoretical predictions \cite{Leitner:2009de} within GiBUU \cite{Buss:2011mx} once (for pions strong) final state interactions have been turned on. This is suprising since the same model describes the photon-induced pion production cross sections on nuclei very well \cite{Krusche:2004uw}.

All these analyses were done in the impulse approximation (IA), where the first interaction happens on one nucleon only. The success of the fits to the QE data with the significantly increased axial mass seems to indicate that changes in the \emph{axial} coupling in the nuclear environment are responsible for the observed excess. On the other hand, a recent model by Bodek et al.\ \cite{Bodek:2011ps,Bodek:2011nk} explains the very same high data in terms of an enhancement of transverse \emph{vector} strength. This model, also in the IA, is the only one so far that also explains the absence of any additional cross section in the energy regime of the NOMAD experiment \cite{Lyubushkin:2008pe}.

Another explanation invokes a very different reaction mechanism. Since processes connected with initial interactions with two nucleons play a role in reactions with electrons on nuclei \cite{Gil:1997bm} it is natural to expect their influence also in neutrino-induced reactions. Indeed, there is now the widespread belief that the $\approx 25\%$ surplus of events labeled as QE in the MiniBooNE experiment over theoretical calculations for true QE scattering is due to such so-called $2p-2h$ events \cite{Nieves:2011yp,Martini:2011wp,Amaro:2011qb}. A short, but comprehensive discussion of the various approaches to describe the $2p-2h$ contributions was recently given by Martini \cite{Martini:2011NF}. Thus, the widely discussed result of a significantly higher axial mass of $M_A \approx 1.2 - 1.3$ GeV extracted from the experiments \cite{AguilarArevalo:2010zc,Gran:2006jn,AguilarArevalo:2010cx} or even $M_A = 1.6$ GeV in \cite{Benhar:2010nx} is then simply a consequence of trying to force a description of $2p-2h$  contributions by a $1p-1h$ model. The latter is also true when treating effects of groundstate correlations  in the  IA \cite{Benhar:2010nx}. The spectral functions used there contain the effects of these correlations in terms of nontrivial connections between energy and momentum of the nucleons. There are, however, many-body contributions, for example (but not only) those connected with exchange currents, that cannot be taken care off by an IA using single particle spectral functions. These extra contributions have to be added explicitly into any theory \emph{and} event generators.

All the theoretical investigations and analyses quoted so far describe also the inclusive electron scattering data in the relevant energy regime and thus pass an important test. To distinguish between the vector and the axial scenarios in the two IA descriptions may be possible by comparing neutrino and antineutrino data. The question then remains how to distinguish between the IA ($1p-1h$ and the $2p-2h$ models. All have aimed at describing the total inclusive cross sections measured by MiniBooNE. While this is clearly a necessary check for any model, it is obviously not sufficient to actually prove the invoked reaction mechanism.
Any experimental verification of the actual reaction mechanism must, therefore, concentrate not just on total, inclusive cross sections, but must instead look at more exclusive channels. Since in the $2p-2h$ and the genuine quasielastic $1p-1h$ processes the number of outgoing nucleons is obviously different the study of knock-out nucleons, numbers and spectra, could possibly help to distinguish between the reaction mechanisms. Data on knock-out nucleons are directly measurable in tracking detectors and do not require the use of a generator for the identification of the reaction process (nucleon knock-out).
Naively one expects the number of knocked-out nucleons to be a direct indication for the relative importance of QE scattering vs.\ other processes, since QE scattering (in vacuo) is connected with only one outgoing nucleon. However, due to nucleon-nucleon and nucleon-pion fsi in a nuclear target  more than one nucleon can be present in the final state even if the initial interaction was with one nucleon only.
\begin{figure*}[tbp]
  \includegraphics[scale=\plotscale]{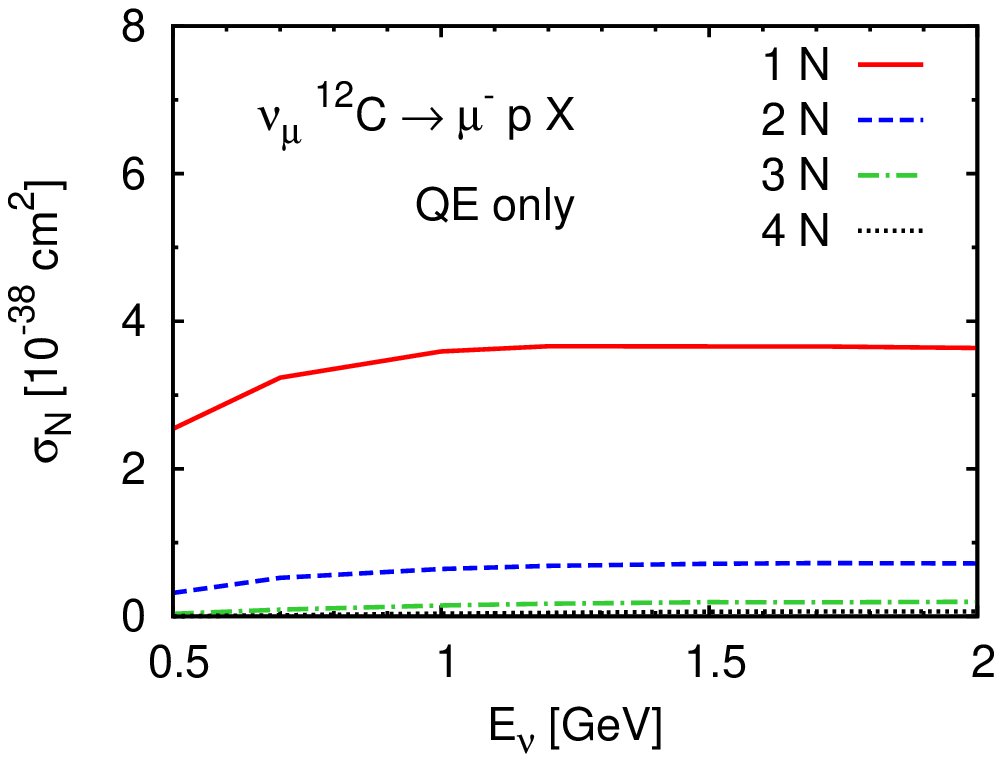}
  \includegraphics[scale=\plotscale]{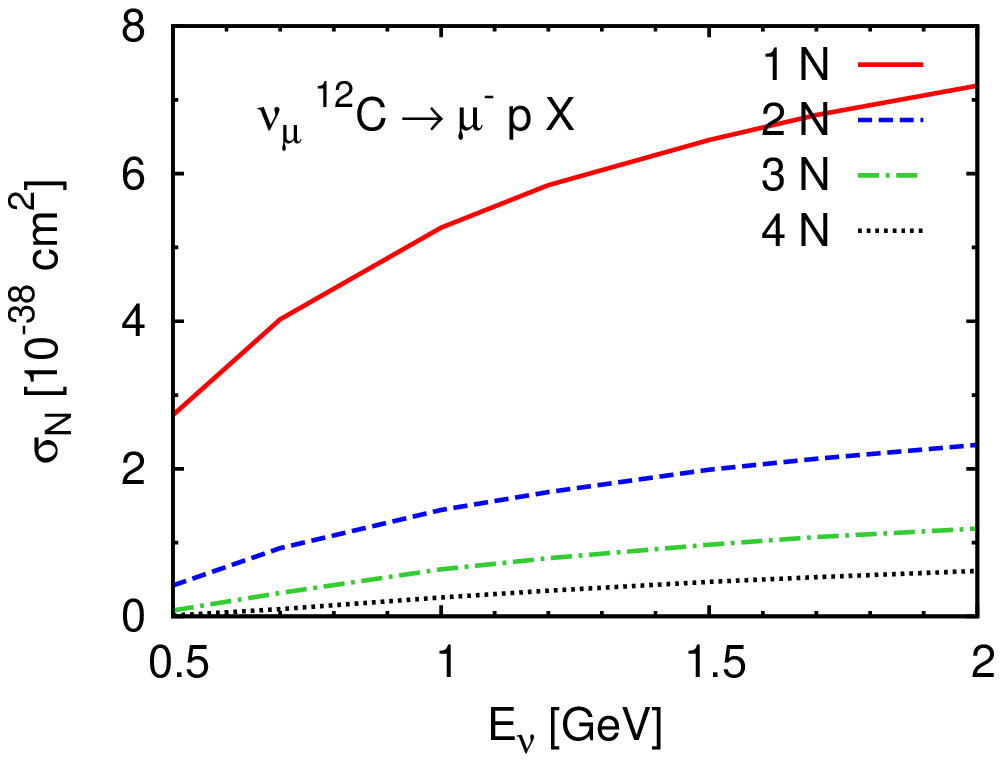}
  \caption{(Color online) Cross section for knock-out of various numbers of nucleons from a $^{12}C$ target as a function of neutrino energy, calculated with GiBUU within the impulse approximation. The picture on the left shows the result obtained from primary QE events only, the right picture contains the result obtained from all primary interactions (from \cite{Leitnerprivcomm}). \label{fig:numberofN}}
\end{figure*}
Fig.\ \ref{fig:numberofN}, based on calculations with GiBUU without any $2p-2h$ interactions \cite{Leitnerprivcomm}, shows this quite well. The figure on the left shows the cross sections for multi-particle knock-out if the initial interaction is only QE scattering (which is not directly observable on a nuclear target). The figure on the right shows the actually observable cross section for $1 \ldots 4$ nucleons. The cross section for seeing more than one nucleon in the final state increases as a function of neutrino energy and the probability to see 2 nucleons amounts to 20\% of the 1-nucleon contribution at 1 GeV even though the initial, first interaction always happened on one nucleon only.

It is to be expected that this number will increase if the $2p-2h$ contributions, that can amount to about 25\% in the total cross section, are taken into account in the generators. At the same time it is to be expected that the energy-spectra of knock-out nucleons are enhanced at lower nucleon energies. These two effects could possibly serve as an experimental verification of the reaction mechanism. The latter is necessary since the many-particle final states will affect the energy reconstruction. We have shown in \cite{Leitner:2010kp} that even the entanglement of initial true quasielastic scattering and $\Delta$ excitation (followed by $\Delta N \to NN$ leads to the appearance of low-energy bumps in the distribution of reconstructed energies. The average uncertainty in the reconstructed energy is about 20\% with a bias towards lower energies; this bias towards lower energies becomes larger with increasing neutrino energy because then pion-production becomes more important. We also have to expect similar effects due to $2p-2h$ processes. In electron-induced reactions the $2p-2h$ excitations fill in mostly the gap between the quasielastic peak and the $\Delta$ peak, i.e.\ they are most effective at higher energy transfers (lower muon energies). If the same is true for neutrinos (there are indications for that in refs.\ \cite{Nieves:2011yp,Martini:2011wp}) it is to be expected that the $2p-2h$ excitations will lead to a shift of the reconstructed energy towards lower energies if the energy reconstruction is done on the basis of quasifree quasielastic scattering kinematics. The latter effect has so far not been investigated, but is under study using GiBUU.

Another question connected with the possible existence of significant two-body processes in the initial interaction is if such processes also contribute to pion production, i.e.\ in processes such as $\nu N N \to N \Delta$. Photoabsorption on nuclei gives an indication for the presence of such effects; the photoabsorption cross section calculated on the basis of a one-nucleon model comes out too low at the upper side of the $\Delta$ resonance (for a discussion and further refs. see \cite{Effenberger:1996im,Carrasco:1989vq}). The large pion cross section observed in experiments with neutrino beams could then also be due to $2p-2h$ contributions. If that is so, this would constitute another complication for the experimental identification of QE scattering, since these additional pion production channels are not being taken out by present-day event generators.

The various reaction mechanisms, such as true QE, pion production, resonance excitation, DIS and $2p-2h$ contributions dominate in different regimes of the energy transfer, and thus also of the neutrino energy. Therefore, any broad-band experiment necessarily contains contributions from the different reaction mechanisms.
 To isolate quasielastic scattering, for example, for use in the neutrino energy reconstruction then requires the use of event \emph{theories \emph{and} generators} which are reliable not only for QE scattering, but are well tested and reliable for a broad class of relevant reactions. In particular, the errors connected with the use of these generators have to be well under control \cite{Harris:2004iq,FernandezMartinez:2010dm,Meloni:2011mr}. The latter can be assessed only by comparison with results from different experiments.
There are lots of data about photo- and electro-production of mesons on nuclear targets, in the energy range of a few 100 MeV (see e.g.\ \cite{Krusche:2004uw}) up to 200 GeV \cite{Ashman:1991cx} which cover the energy regime relevant for present day's neutrino experiments. These reactions involve a closely related primary interaction and the very same fsi as in the neutrino-induced reactions and can thus be used to check the reliability of the generators. So far, only GiBUU \cite{Buss:2011mx} has undergone this 'stress test'.

\paragraph{Summary}
The excess observed in the QE scattering cross sections of neutrinos on nuclear targets has been described both within the impulse-approximation or within a two-nucleon interaction picture. Within the former framework the data have been fitted by invoking either an increase in axial or in vector strength. Connected with this `quasielastic-like' excess may be the high pion production yield. To unambiguously identify the reaction mechanism more exclusive experiments will have to be performed and the $2p-2h$ interactions must be incorporated into event generators. Nucleon knock-out represents the most promising reaction to study. Comparison between neutrino and antineutrino results may help to distinguish between the axial and the vector scenarios.

\begin{theacknowledgments}
I am grateful to Murat Kaskulov, Olga Lalakulich and Tina Leitner for many helpful discussions on neutrino-nucleus interactions. This work was supported by Deutsche Forschungsgemeinschaft (DFG) and HIC for FAIR.
\end{theacknowledgments}

\end{document}